\documentclass[amssymb,amsmath,twocolumn,floatfix,apl]{revtex4}
\usepackage{graphicx}
\usepackage[usenames]{color}

\pagecolor{white}
\begin{document}
\color{black}
\newcommand {\ket}[1]{\ensuremath{| #1 \rangle}}
\newcommand {\bra}[1]{\ensuremath{\langle #1 |}}
\newcommand {\vecon}{\ensuremath{\dot{\times}}}

\title{A multiplexed single electron transistor for application in scalable solid-state quantum computing.}
\date{\today}
\author{Vincent I. Conrad\footnote{Author to whom correspondence should be addressed.}, Andrew D. Greentree, and Lloyd C. L. Hollenberg}
\affiliation{Centre for Quantum Computer Technology, School of
Physics, University of Melbourne, Vic. 3010, Australia}

\begin{abstract}
Single Electron Transistors (SETs) are nanoscale electrometers of
unprecedented sensitivity, and as such have been proposed as
read-out devices in a number of quantum computer architectures. We
show that the functionality of a standard SET can be multiplexed so
as to operate as both read-out device \emph{and} control gate for a solid-state qubit. Multiplexing in this way may be critical in lowering overall gate densities in scalable
quantum computer architectures. 
\end{abstract}

\maketitle

The Single Electron Transistor (SET) is a device that can act as an
exquisitely sensitive electrometer.  This sensitivity derives from
precise control of the absolute charge state of a small, metallic
island, coupled via tunnel junctions to macroscopic leads.  Since
Fulton and Dolan's \cite{bib:FultonPRL1987} initial experiments, SETs
have been suggested for a diverse range of applications, from elements for
classical logic \cite{bib:UchidaIEEE2003} to single-photon detectors\cite{bib:KomiyamaNature2000}.  SETs are commonly suggested as readout mechanisms for quantum scale devices\cite{bib:ColePRB2005} and quantum computers (QC) \cite{bib:KaneNature1998}, either via direct sensing of charge qubits \cite{bib:Aassime2001,bib:GreentreePRB2004,bib:StacePRB2004}, or of spin qubits after an initial spin to charge transduction process \cite{bib:KaneNature1998,bib:HollenbergPRB2004,bib:GreentreePRB2005,bib:IonicioiuNJP2005,bib:Testolin2005}.  Compatibility of SETs with such rigorous demands has been shown often, see for
example Refs.~\onlinecite{bib:DevoretNature2000,bib:BuehlerAPL2005}.
Reviews of SETs can be found in (for example)
Refs.~\onlinecite{Esteve_1992,bib:HollenbergReview}.

Merely showing SETs have the required sensitivity for qubit readout
is not, however, sufficient for the development of a scalable
quantum computer architecture.  Of principle concern in this paper
is the requirement for minimal gate density in the surface metal
layer \cite{bib:GateDensity}.  Standard designs for SETs usually
have a relatively large footprint ($\gtrsim 10^4 \mathrm{nm}^2$\cite{bib:BuehlerJAP2004}), which with attendant
control gates may be problematic in terms of spacing in the
original Kane 1-D \cite{bib:KaneNature1998} and the scalable 2-D
\cite{bib:Hollenberg2D} QC architectures. Antenna structures
\cite{bib:Clark,bib:LeeNano2005,Conrad_2005} may be of some
assistance in packing in all the required elements, however, as the
number of readout elements is increased from proof of principle
devices to fully operational QCs, we believe a degree of multiplexed
functionality will be very advantageous.  

\begin{figure}[tb!]
\centering
\includegraphics[width=0.4\textwidth]{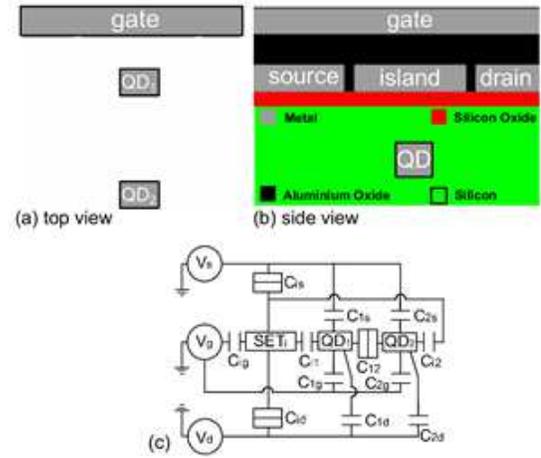}
\caption{(Color online) (a) \& (b) Schematic of SET designed for minimizing gate density in a scalable QC.  (c) Circuit diagram for the Single Electron Transistor coupled to two Quantum Dots (SET-2QD) system.  The SET consists of three continuously variable voltage sources (source, drain, and gate), coupled to an isolated island.  The source and drain are connected to the island by tunnel junctions, allowing for current flow through the SET.  Isolated regions in which effects due to electron occupation number are critical have been indicated with boxes.
} 
\label{fg:setup} 
\noindent
\end{figure}

In this paper, we
show that a SET can be used for both qubit control and
read-out, reducing the number of electrodes required for operation
of the QC.  We also propose a `strip' design SET, with the gate placed above the source and drain [Fig. \ref{fg:setup} (a) and (b)] to further reduce the density of surface electronics.  We consider a charge-based, double quantum dot (QD) qubit in our model, as read-out of a spin qubit may follow the same procedure, via spin to charge transduction.  The qubit states correspond to the localization of a shared electron between one of two QDs.  The circuit for our model is shown in Fig. \ref{fg:setup}.
Read-out via a SET consists of inferring the state of the 
qubit by the current through the SET. The current depends on the
state of the qubit, and also on the potential differences between
the SET island and any other electrodes capacitively coupled
to it.  By carefully controlling the potential landscape of the entire system, we show
that the current through a SET can be kept constant, ideally zero,
while the potentials on the leads vary relative to the circuit
ground. Hence the potential on the qubit can be made to
vary, allowing the SET to act as a qubit control gate.

We work in the steady state regime where the current through the SET is modeled via energy minimization arguments of the entire
system, following the orthodox theory of single electron
tunneling \cite{Esteve_1992,Conrad_2005}.  We begin our discussion by presenting data from a SET model without the QDs.  We then show results of the SET-2QD system and present protocols for both control and read-out modes.

\begin{figure}[httb!]
\centering
\includegraphics[width=0.4\textwidth]{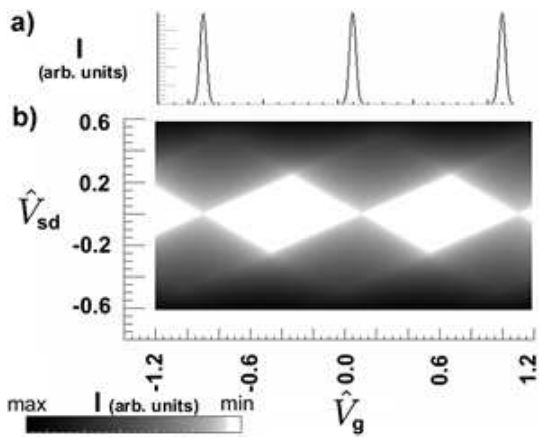}
\caption{(a) Current through a SET vs dimensionless gate voltage ($\hat{V_g}=V_g C_{ig}/e$) and source-drain potential difference ($\hat{V_{sd}}=V_{sd}(C_{is}+C_{id})/e$) for the device geometry of Fig.~\ref{fg:setup} without the coupled qubit.  Maxima occur when there is degeneracy of neighboring charge occupations, in between
the device is `blockaded' and sequential charge tunneling is
suppressed.  (b) Spanning the source-drain bias space transforms the Coulomb blockade regions into the familiar Coulomb diamonds.}
\label{fg:coulomb_blockade} \noindent
\end{figure}

Following Ref.~\onlinecite{Conrad_2005}, we determine the current
through a SET, ignoring cotunneling processes. Coulomb blockades are shown in Fig.~\ref{fg:coulomb_blockade} (a).  We  gain further insight into the SET behavior by studying the current as a function of both the potential difference between the
source and drain electrodes, and the gate potential as shown in Fig. \ref{fg:coulomb_blockade}(b), where we have normalized the input voltages: $\hat{V_g}=V_g C_{ig}/e$, $\hat{V}_{sd}=V_{sd}(C_{is}+C_{id})/e$, where e is the magnitude of the charge of an electron.  It is clear from Fig. \ref{fg:coulomb_blockade}(b) that the current through a SET is not uniquely determined by the electrode voltages. Therefore, there is some freedom to choose the potentials on the SET
source, drain and gate electrodes to obtain a particular current
through the SET.  It is this fact that allows the use of a SET as a qubit
control gate.

We now describe our model for the SET-2QD system, and present the method by which the SET can be used for both control and read-out of the qubit.  The behavior of mesoscopic circuitry is dependent on the interplay
between the continuous variable given by the definition of
capacitance, $C \equiv Q/V$, and the actual number of excess
electrons on an isolated region of the circuit (SET island and QDs).  We refer to the continuous
variable as the virtual charge.  The virtual charge induced on any isolated region is $\tilde{Q}_\alpha = \sum_\beta C_{\alpha \beta} V_{\beta}$.
The full vector $\tilde{\mathbf{Q}}$, describing the virtual charge induced on each isolated
region of the system is then $\tilde{\mathbf{Q}}=C_c\mathbf{V}$, where $C_c$ is the correlation capacitance matrix describing
capacitive coupling between the electrodes and quantized charge
regions, and $\mathbf{V}$ is the vector of electrode potentials.  Labeling each QD with a numeric
subscript we write

\begin{equation}
\label{eq:Cc_and_V} C_c = \left[\begin{array}{ccc} C_{is} & C_{id} &
C_{ig}
\\C_{1s} & C_{1d} & C_{1g}
\\C_{2s} & C_{2d} & C_{2g}
\end{array}\right],
\mathbf{V}=
\left[\begin{array}{ccc}
V_{s} 
\\V_{d}
\\V_{g}
\end{array}\right] ,
\end{equation}
The charging energy of the entire SET-2QD system is determined by,
\begin{equation}
\label{eq:CircuitEnergy}
E=\frac{1}{2}\mathbf{Q}^{\mathrm{T}}C^{-1}_E\mathbf{Q}\;,
\end{equation}
where $\mathbf{Q}$ is a vector containing the virtual and actual charge on all
isolated sections of the circuit (i.e. the island and QDs), and
$C_E$ is the `energy' capacitance matrix (defined below).  The total
charge is expressed as $\mathbf{Q}=\tilde{\mathbf{Q}} - \mathbf{n}e$,
where $\mathbf{n}$ is a vector containing the excess electrons on
each isolated region. $C_E$ describes the cross capacitances of the QDs and the
island and is given by
\begin{equation}
C_{E}= \left[\begin{array}{ccc} C_{i\Sigma} & -C_{i1} & -C_{i2}
\\-C_{1i} & C_{1\Sigma} & -C_{12}
\\ -C_{21} & -C_{2i} & C_{2\Sigma}
\end{array}\right]\;,
\end{equation}
where $C_{\alpha\Sigma}$ is the total capacitance of the $\alpha$ isolated region to all other objects in the system ($C_{\alpha \Sigma}= \sum_{m\neq\alpha} C_{\alpha m} $).
The above relations allow the determination of the current through
the SET for the full SET-2QD system for any voltage on the
electrodes.  We calculate the current based on a standard master equation approach \cite{Conrad_2005} using the rate of change of the number of electrons on the island.  The rate at which electrons tunnel from the island to the drain, given $n$ electrons occupy the island is given by, 
\begin{equation}
\Gamma_{di}^n=
\frac{1}{q_{e}^{2}R_{t}}
\frac{\Delta E_{di}^n}{\exp\left (\frac{\Delta E_{\chi i}^n}{k_{B}T} \right )-1}.  
\end{equation}
Where $\Delta E$ is given by the energy of the system before and after the event based on Eq. \ref{eq:CircuitEnergy}.  Similar expressions apply for electrons tunneling from the drain or to/from the source.  The resistance is calculated based on the materials and geometries of the SET using
\begin{equation}
R_{t}=\frac{\hbar^3\exp(\frac{2W}{\hbar}\sqrt{2m_e\phi})}{2\pi m_e^* q_{e}^2 E_F A}, 
\end{equation}
where W is the width of the tunnel barrier and A is its surface area.  The variable $\phi$ is the height of the potential barrier (taken to be a typical value of 2eV\cite{Koppinen_2003} and $E_F$ is the Fermi energy of the island (taken to be 11.65eV at 4K\cite{Brenner_2004}).  We use an effective mass of an electron in aluminium oxide of $m_e^*=0.35m_e$\cite{Xu_1991}.  The present limit to detectable SET currents is in the fA regime\cite{bib:Bylander2005}.  We choose to scale the geometry of Fig. \ref{fg:setup} (a) to produce a minimal footprint such that the electrodes' height and width are 10nm, and the tunnel junctions are 3nm, while remaining above this limit.  Reducing the resistance will provide a commensurate increase in current without effecting the conclusions of this paper.

\begin{figure}[t!]
\centering
\includegraphics[width=0.5\textwidth]{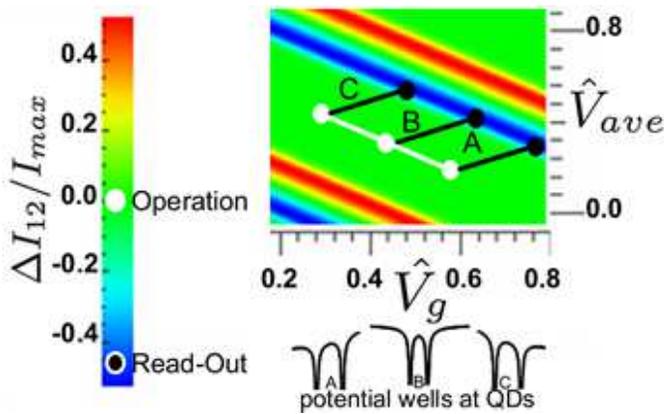}
\caption{(Color online) Difference in current through SET for a single electron occupying QD1 or QD2 ($\Delta I_{12}$) in Fig.\ref{fg:setup}.  Voltages are normalised: $\hat{V_g}=V_g C_{ig}/e$, $\hat{V}_{ave}=(V_s+V_d)(C_{is}+C_{id})/2e$.  The difference in current for the two states of the qubit allows for read-out at points of maximum current variation (e.g. black dots).  Regions with no current variation for the qubit states can be used for qubit operations.  Single qubit rotations can occur via time evolution at a point in this region (white dots) or by varying the potential landscape near the QDs (white line).  Paths along A B and C (or any other parallel path) maintain a constant potential difference between the QDs (as shown in corresponding potential well diagrams), allowing the SET to be multiplexed for both read-out and operation.} 
\label{fg:VbarvsVg} \noindent
\end{figure}

Fig.~\ref{fg:VbarvsVg} displays the current variation through the SET for the qubit electron being localized on each of the QDs.  We present the current variation as a
function of the the normalized gate voltage ($\hat{V}_g$) and the normalized average of the potentials on the source and drain $\hat{V}_{ave}=(V_s+V_d)(C_{is}+C_{id})/2e$ (maintaining a constant source-drain bias).  By maintaining constant current, (ideally zero) through the SET, but varying the SET electrode's
potentials, one can remain in a blockaded region of
Fig.~\ref{fg:VbarvsVg} whilst changing the local electrostatic
potential landscape.  The paths A, B, and C in Fig. \ref{fg:VbarvsVg} (and any other parallel path) produce a constant potential difference between the QDs (as depicted in the corresponding potential well diagrams in Fig. \ref{fg:VbarvsVg}).  These paths were calculated using Eq. \ref{eq:CircuitEnergy}, as a constant potential difference between the QDs is equivalent to a constant difference in induced charge.  Along the white line in Fig. \ref{fg:VbarvsVg} the SET has no current passing through it, effectively turning off the measurement of the qubit.  Single qubit operations may therefore be performed by traversing the white line in Fig. \ref{fg:VbarvsVg} or by time evolution at a point (white dots).  Read-out would be performed at points of maximum current variation (e.g. black dots).  Thus, appropriately traversing the voltage space of the system allows the SET to be multiplexed for both control and read-out. 

We have shown that a full exploration of the parameter space for SET
operation allows a SET to be operated as either a read-out device
\emph{or} control gate independently.  Such
multiplexed functionality allows for a significant reduction in overall gate
density, which we believe to be necessary for practical quantum
computing. We have also presented a `strip' design SET to further reduce gate densities. 

This work was supported by the Australian Research Council, the Australian Government, the US National Security Agency (NSA), Advanced Research and Development Activity (ARDA), and the Army Research Office (ARO) under contract number W911NF-04-1- 0290.


 \end{document}